\documentclass[
prd
 ,secnumarabic%
,amssymb,nobibnotes, aps, prd]{revtex4}
\usepackage{amsmath}
\usepackage{graphicx}
\input{epsf}

\newcommand{\be}{\begin{equation}}    
\newcommand{\ee}{\end{equation}}
\newcommand{\beq}{\begin{eqnarray}}
\newcommand{\eeq}{\end{eqnarray}}
\newcommand{\beqn}{\begin{eqnarray*}}
\newcommand{\eeqn}{\end{eqnarray*}}
        
\newcommand{\p}{\partial}

\def\nn{\nonumber}

\def\ver{\vskip 12pt}

\def\IL{\relax{\rm I\kern-.18em L}}

\begin{document}


\draft

\title{Quasinormal modes of Reissner-Nordstr\"om-anti-de Sitter black holes:
scalar, electromagnetic and gravitational perturbations}

\author
{E. Berti and K.D. Kokkotas}
\address
{Department of Physics, Aristotle University of Thessaloniki,
Thessaloniki 54124, Greece}

\date{\today}

\begin{abstract}

We study scalar, electromagnetic and gravitational perturbations of a
Reissner-Nordstr\"om-anti-de Sitter (RN-AdS) spacetime, and compute
its quasinormal modes (QNM's). We confirm and extend results
previously found for Schwarzschild-anti-de Sitter (S-AdS) black holes.
For ``large'' black holes, whose horizon is much larger than the AdS
radius, different classes of perturbations are almost exactly {\it
isospectral}; this isospectrality is broken when the black hole's
horizon radius is comparable to the AdS radius. We provide very
accurate fitting formulas for the QNM's, which are valid for large
black holes having charge $Q<Q_{ext}/3$ ($Q_{ext}$ being the
extremal value of the charge).

Electromagnetic and axial perturbations of large black holes are
characterized by the existence of pure-imaginary (purely damped)
modes. The damping of these modes tends to infinity as the black hole
charge approaches the extremal value; if the corresponding mode
amplitude does not tend to zero in the same limit, this implies that
{\it extremally charged RN-AdS black holes are marginally unstable}.
This result is relevant in view of the AdS/CFT conjecture, since,
according to it, the AdS QNM's give the timescales for approach to
equilibrium in the corresponding conformal field theory.
\end{abstract}

\pacs{PACS numbers: 04.70.Bw, 04.50.+h}

\maketitle

\narrowtext

\section{Introduction}

\ver\ver 

Quasinormal modes (QNM's) are well known to play an important role in
black hole physics. They determine the late-time evolution of fields
in the black hole exterior, and numerical simulations of stellar
collapse and black hole collisions have shown that in the final stage
of such processes (``ringdown'') the quasinormal modes eventually
dominate the black hole response to any kind of perturbation. For
these reasons, QNM's of black holes in asymptotically flat spacetime
have been extensively studied for more than thirty years (for
comprehensive reviews see \cite{KS,N}).

Perturbations of non-asymptotically flat spacetimes have also been
studied for years in a cosmological context.  For example,
perturbations of Schwarzschild-de Sitter spacetimes were considered in
\cite{GN}, and the corresponding QNM's were computed in \cite{OF}. An
analogous study for Reissner-Nordstr\"om-de Sitter perturbations was
performed in \cite{MM}.

In the last few years, the anti-de Sitter (AdS) conformal field theory
(CFT) conjecture has led to an intensive investigation of black hole
QNM's in asymptotically AdS spacetimes. According to the AdS/CFT
correspondence \cite{M}, a large static black hole in asymptotically
AdS spacetime corresponds to an (approximately) thermal state in the
CFT. So the time scale for the decay of the black hole perturbation,
which is given by the imaginary part of its QNM's, corresponds to the
timescale to reach thermal equilibrium in the strongly coupled CFT
\cite{BSS}. The computation of these time scales in the CFT is a
difficult task. Therefore, black hole perturbation theory has turned
out to be, quite unexpectedly, a useful tool to compute CFT
thermalization time scales.

Quasinormal modes in AdS spacetime were first computed for a
conformally invariant scalar field, whose asymptotic behaviour is
similar to flat spacetime, in \cite{CM}.  Subsequently, motivated by
the AdS-CFT correspondence, Horowitz and Hubeny made a systematic
computation of QNM's for scalar perturbations of Schwarzschild-AdS
(S-AdS) spacetimes \cite{HH}.  Their work was extended to
gravitational and electromagnetic perturbations of S-AdS black holes
in \cite{CL}. A reconsideration of the correct boundary conditions for
the QNM problem was presented by Moss and Norman \cite{MN}, who
considered gravitational perturbations of both Schwarzschild-de Sitter
and S-AdS spacetimes.  Finally, the study of scalar perturbations was
extended to the case of Reissner-Nordstr\"om-AdS (RN-AdS) black holes
in \cite{WLA}.

Frequency-domain studies of de Sitter and anti-de Sitter black hole
spacetimes have lately been complemented by time evolutions.  In
\cite{BCKL} and \cite{BCLP}, the authors used a time evolution
approach to study tails in Schwarzschild-de Sitter backgrounds from a
cosmological perspective.  Motivated by the AdS/CFT correspondence,
analogous time evolutions have recently been performed for RN-AdS
backgrounds in \cite{WMA}.

Numerical methods used so far to compute QNM's in AdS backgrounds tend
to break down, in the S-AdS case, as the black hole horizon becomes
much smaller than the AdS radius \cite{HH}. Therefore, special
attention has been devoted to the small-black hole limit, using both a frequency domain  \cite{K} and a time domain approach \cite{ZWA}.

Our aim in this paper is to present a comprehensive study of scalar,
electromagnetic and gravitational perturbations of RN-AdS black holes
in the frequency domain. We verify (or sometimes disprove) and extend
many results obtained in the aforementioned papers. We also compute
for the first time electromagnetic and gravitational QNM's for RN-AdS
black holes, building on work done previously in the cosmological
context for Reissner-Nordstr\"om-de Sitter spacetimes \cite{MM}.

The plan of the paper is as follows. In section \ref{eqns} we
introduce our notation for the background metric, display explicitly
the wave equations describing scalar, electromagnetic and
gravitational perturbations, and briefly present our numerical methods
and conventions. Details on the numerical methods are deferred to the
Appendix. In section \ref{numres} we first summarize and extend some
well-known results for perturbations of S-AdS black holes, considering
all three cases (scalar, electromagnetic and gravitational
perturbations); then we use the uncharged case as a a starting point
to discuss perturbations of RN-AdS black holes, and present most of
our new numerical results. The conclusions and a discussion follow.

\section{Perturbation equations in a RN-AdS background}
\label{eqns}
\subsection{Background}

We are interested in scalar, electromagnetic and gravitational
perturbations of a Reissner-Nordstr\"om anti-de Sitter metric,
\be
ds^2=f(r)dt^2-{dr^2\over f(r)}
-r^2(d\theta^2+\sin^2\theta d\phi^2),
\ee
where $f(r)=\Delta/r^2$ and
\be
\Delta=r^2-2Mr+Q^2+{r^4\over R^2}.
\ee
In the previous formulas $M$ is the black hole mass, $Q$ the black
hole charge, and $R$ is the AdS radius.
The black hole mass is related to its charge $Q$ and horizon radius
$r_+$ by the relation
\be
M={1\over 2}\left(r_++{r_+^3\over R^2}+{Q^2\over r_+}\right),
\ee
while the black hole's Hawking temperature is given by
\be
T={1-Q^2/r_+^2+3r_+^2/R^2\over 4\pi r_+}.
\ee
It will be useful in the following to introduce a tortoise coordinate
$r_*$, defined in the usual way by the relation
\be
{d r\over d r_*}={\Delta\over r^2}.
\ee
The extremal value of the black hole charge, $Q_{ext}$, is given by
the following function of the hole's horizon radius:
\be
Q_{ext}^2=r_+^2\left(1+{3r_+^2/R^2}\right).
\ee

\subsection{Scalar perturbations}

The scalar perturbations of the RN-AdS metric have been studied in
\cite{WLA}, so we only give the final result. Separating the angular
dependence of the perturbations and introducing a differential
operator
\be
\Lambda^2={d^2\over dr_*^2}+\omega^2,
\ee
the radial part of the perturbations obeys a wave equation
\be\label{scalar}
\Lambda^2Z=V Z
\ee
with potential
\beq
V&=&f(r)\left[{l(l+1)\over r^2}+{f'(r)\over r}\right]\\
&=&
{\Delta\over r^2}\left[{l(l+1)\over r^2}
+{1\over r}\left(
{r_+\over r^2}+{r_+^3\over R^2r^2}+{Q^2\over r_+r^2}
-{2Q^2\over r^3}+{2r\over R^2}
\right)\right].
\nn
\eeq
Here $l$ is the usual spherical harmonic index.
In the limit $Q=0$ this potential reduces to the four-dimensional
limit of the scalar potential considered by Horowitz and Hubeny
\cite{HH}.

\subsection{Electromagnetic and gravitational perturbations}

The electromagnetic and gravitational perturbation equations for
Reissner-Norstr\"om-de Sitter black holes were considered in \cite{MM}
following the track outlined in \cite{C}. Once again, their derivation
is carried over with no difficulty to the RN-AdS spacetime, so we only
give the final expressions for the perturbation equations.  After
separation of the angular dependence, the axial perturbations (which we
will denote by a superscript ``-'') are described by a couple of wave
equations:
\be\label{axial}
\Lambda^2Z_i^-=V_i^-Z_i^-.
\ee
Defining
\be\label{defn}
n={(l-1)(l+2)\over 2}
\ee
and
\beq
p_1&=&3M+(9M^2+8nQ^2)^{1/2},\\
p_2&=&3M-(9M^2+8nQ^2)^{1/2},\nn
\eeq
the relevant potentials are
\be\label{axpot}
V_i^-={\Delta\over r^5}
\left[
2(n+1)r-p_j\left(1+{p_i\over 2nr}\right)
\right]
\qquad(i,j=1,2; i\neq j).
\ee
Similarly, the radial parts of the polar perturbations (that we will
denote by a superscript ``+'') obey the couple of equations:
\be
\Lambda^2Z_i^+=V_i^+Z_i^+,
\ee
where
\beq\label{polar}
V_1^+&=&{\Delta\over r^5}\left[U+{1\over2}(p_1-p_2)W\right],\\
V_2^+&=&{\Delta\over r^5}\left[U-{1\over2}(p_1-p_2)W\right].\nn
\eeq
Here $U$ and $W$ are given by
\beq
W&=&{\Delta\over r\varpi^2}\left(2nr+3M\right)+{1\over \varpi}
\left(nr+M-{2r^3\over R^2}\right),\\
U&=&\left(2nr+3M\right)W+
\left(\varpi-nr-M+{2r^3\over R^2}\right)-{2n\Delta\over \varpi},\nn
\eeq
and the function $\varpi$ is defined by the relation
\be
\varpi=nr+3M-{2Q^2\over r}.
\ee

In the limit $Q=0$, the potentials $V_1^-$ and $V_1^+$ reduce to the
corresponding potential for pure electromagnetic perturbations of
Schwarzschild-AdS black holes, equation (5) in \cite{CL}:
\be
V_{EM}=\Delta {l(l+1)\over r^4}.
\ee

Furthermore, in the same limit, the potential $V_2^-$ reduces to 
the potential for pure axial gravitational perturbations of Schwarzschild,
equation (13) in \cite{CL}:
\be\label{Vodd}
V_{odd}={\Delta\over r^2}\left[{l(l+1)\over r^2}-{6M\over r^3}\right],
\ee
and the $V_2^+$ potential reduces to the potential describing pure
polar gravitational perturbations of Schwarzschild, equation (17) in
\cite{CL}:
\be
V_{even}=\left({2\Delta\over r^5}\right){
9M^3+3n^2Mr^2+n^2(1+n)r^3+3M^2(3nr+3r^3/R^2)\over
(3M+nr)^2}.
\ee

\subsection{Numerical method}

In AdS backgrounds, QNM's are defined as solutions to the relevant
wave equations characterized by purely ingoing waves at the event
horizon and vanishing perturbation at radial infinity.  We used two
different numerical methods to solve this eigenvalue problem. The
first has been discussed at length in previous papers
\cite{HH,CL,WLA}, and we briefly recall it using our notation (and
covering the slightly more general case considered here) in Appendix
\ref{nummeth}. As a further check, we also extended to the axial
perturbations of RN-AdS black holes the Fr\"obenius method used in
\cite{MN} for S-AdS black holes, developing on previous work for black
holes in asymptotically flat spacetime \cite{L,L2}. This extension
(which essentially involves the use of a four-term recurrence relation
instead of a three-term recurrence relation) is presented in Appendix
\ref{nummeth2}.

The two methods yield numerical results which agree to machine
accuracy.  In both formulations of the eigenvalue problem, we
numerically search for roots in the plane $(\omega_R,\omega_I)$ using
M\"uller's method.  For pure-imaginary roots, it is sometimes
convenient to resort to Ridder's method (for a description of both
root-searching techniques, see e.g. \cite{NR}). The Fr\"obenius method
is computationally much faster, and it is also easier to
implement. However, at variance with the asympotically flat case,
axial and polar perturbations in anti-de Sitter backgrounds are not
isospectral \cite{CL}. A similar isospectrality breaking was found for
perturbations of charged, dilaton black holes in \cite{FPP}.  As a
consequence, polar perturbations, which are described by the rather
messy potentials (\ref{polar}), cannot simply be related to axial
perturbations through a Chandrasekhar transformation of the radial
perturbation variables \cite{C}. So we only used the Fr\"obenius
method as a check when computing modes associated to the axial
potentials (\ref{axpot}).

From now on, in presenting the numerical results, we will set the AdS
radius $R=1$.  It is also useful to normalize the charge to its
extremal value, defining
\be
\bar Q=Q/Q_{ext}.
\ee
Finally, we set $\omega=\omega_R-i\omega_I$, in order to have positive
values for the imaginary parts of QNM frequencies.  In the next
sections we will first recall the properties of S-AdS QNM's, verifying
and extending results found in \cite{HH,CL}.  Then we will present our
results for charged black holes, discussing both the scalar modes
studied in \cite{WLA} and the electromagnetic and gravitational
perturbations, which, to our knowledge, are studied here for the first
time. In both cases we will first discuss the ``ordinary'' QNM's, and
then the purely damped QNM's (whose real part $\omega_R=0$), which
show rather peculiar features deserving special attention.

\section{Numerical results}\label{numres}

\subsection{Schwarzschild-anti-de Sitter black holes}

\subsubsection{``Ordinary'' quasi-normal modes}

\begin{figure}[h]
\centering
\includegraphics[angle=270,width=8.5cm,clip]{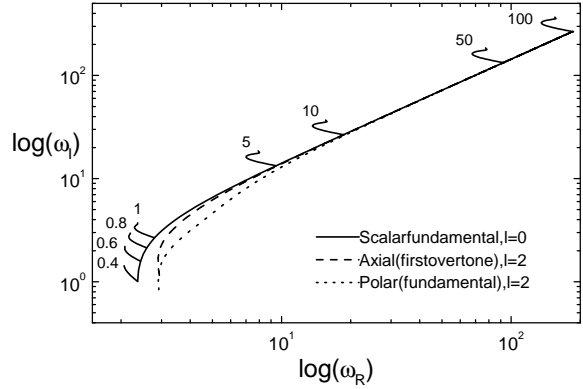}
\caption{
Plot of the S-AdS QNM frequencies for the fundamental scalar mode with
$l=0$ (continuous line), for the first non-purely damped axial mode
with $l=2$ (dashed line) and for the fundamental polar mode for $l=2$
(dotted line). All calculations are started at $r_+=100$ (top right of
the diagram). The step in $r_+$ is reduced by $\Delta r_+=0.1$ and the
calculation is iterated until our numerical code fails to converge
(bottom left). Corresponding to selected values of the horizon radius
(namely, $r_+=100,~50,~10,~5,~1,~0.8,~0.6,~0.4$, whose S-AdS
frequencies are tabulated in Table I of Horowitz and Hubeny), we
``switch on'' the charge and follow the modes in the complex plane.
The resulting trajectories are the ``tails'' departing from the
continuous scalar QNM line. Axial and polar perturbations are almost
isospectral both in the large and in the small black hole limit.
}
\label{fig1}
\end{figure}

Consider a sequence of {\it uncharged} black holes in anti-de Sitter
space, and fix units such that the AdS radius $R=1$. Then QNM
frequencies only depend on the holes' horizon radius $r_+$, and can be
plotted as curves in the complex $(\omega_R,~\omega_I)$ plane. Such
curves are shown for the fundamental scalar mode with $l=0$
(continuous line), for the first non-purely damped axial mode with
$l=2$ (dashed line) and for the fundamental polar mode with $l=2$
(dotted line) in figure \ref{fig1}. Comparing the dashed and dotted
lines in the figure, it is apparent that, as mentioned in the previous
paragraph, in a S-AdS background the isospectrality between even and
odd perturbations (which is a property of asymptotically flat black
hole spacetime perturbations) is broken. However, it is approximately
restored in the large $\omega$ and in the small black hole limit (for
a discussion of this property, see section III C in \cite{CL}). Scalar
perturbations are isospectral with odd and even parity gravitational
perturbations in the large black hole (large $\omega$) limit. As yet,
there is no analytical proof of this rather surprising feature, which
we have numerically verified both for the fundamental mode and for the
overtones.

In general, our results are in perfect numerical agreement with those
presented in \cite{HH,CL}.  In their treatment of {\it scalar}
perturbations, Horowitz and Hubeny found that, for large black holes
and $l=0$, the real and imaginary part of QNM frequencies scale
linearly with the temperature. In four dimensions, they predicted that
the fundamental mode frequencies are well fitted by the relations
$\omega_R=7.75T$, $\omega_I=11.16T$.  They also found that, in
general, the overtones are equally spaced, obeying relations of the
form $\omega_R(n)\sim 54+131n$, $\omega_I(n)\sim 41+225n$, where $n$
is a positive integer, not to be confused with the coefficient $n$
defined in equation (\ref{defn}).  As $l$ increases, $\omega_I$
decreases and $\omega_R$ increases, but the $l$-dependence of the
modes is extremely weak.

\begin{table}
\centering
\caption{
Linear scaling of the real and imaginary part of QNM frequencies with
the black hole temperature for large S-AdS black holes. All fits have
been obtained on a set of numerical data with black hole radii varying
between $r_+=10$ and $r_+=100$ (see text). In parentheses, we show the
difference between a given fitting coefficient and the fitting
coefficient of the preceding mode: as the mode order grows, modes tend
to become more and more equally spaced. The axial modes have been
labelled starting from one to take into account the existence of a
purely damped mode.
}
\vskip 12pt
\begin{tabular}{@{}ccccccccc@{}}
\multicolumn{3}{c}{Scalar ($l=0$)} 
&\multicolumn{3}{c}{Axial ($l=2$)} 
&\multicolumn{3}{c}{Polar ($l=2$)} \\
\hline
Mode &$\omega_R/T$ &$\omega_I/T$ 
&Mode &$\omega_R/T$ &$\omega_I/T$ 
&Mode &$\omega_R/T$ &$\omega_I/T$ 
\\
\hline
0  & 7.747   & 11.157 & 1 & 7.748  & 11.157 & 0 & 7.750 & 11.153 \\
\hline		                               			
1  & 13.243  & 20.592 & 2 & 13.243 & 20.592 & 1 & 13.246  & 20.585\\
   &(5.495)  &(9.435) &   &(5.495) &(9.435) &   &(5.496)  &(9.432)\\
\hline
2  & 18.702  & 30.021 & 3 & 18.702 & 30.021 & 2 & 18.705  & 30.010\\
   &(5.458)  &(9.429) &   &(5.459) &(9.429) &   &(5.459)  &(9.425)\\
\hline
3  & 24.151  & 39.447 & 4 & 24.152 & 39.447 & 3 & 24.154  & 39.434\\
   &(5.450)  &(9.426) &   &(5.450) &(9.426) &   &(5.449)  &(9.424)\\
\end{tabular}
\label{fitovertones}
\end{table}

Our numerical results confirm all of these statements, and show that
they apply also to axial and polar gravitational perturbations.  Table
\ref{fitovertones} shows the fit parameters for the linear
temperature-dependence of the lowest QNM frequencies with $l=0$
(scalar case) and $l=2$ (gravitational case). Notice that the
labelling of axial QNM frequencies starts at $n=1$, since we regard
the pure imaginary mode found in \cite{CL} as the fundamental. All
fits have been done on a table obtained performing explicit
calculations for black holes with horizon radii varying from $r_+=10$
to $r_+=100$ in steps of $\Delta r_+=0.1$. Values in parentheses
indicate the spacing in the fitting coefficient between a given mode
and the previous overtone: as the overtone number increase, these
numbers tend to a constant. It may also be noted that fitting
coefficients in this large black hole limit are essentially the same
for all three kinds of perturbations.

In brief, from Table \ref{fitovertones} we can deduce a very simple
``generalized formula'': for the fundamental mode and the overtones of
{\it scalar and gravitational} S-AdS perturbations, QNM frequencies
are well approximated by the fitting formula
\be\label{linear2} 
\omega_R\sim (7.75+5.46n)T,\qquad
\omega_I\sim (11.16+9.43n)T.
\ee
A relabelling of the integer $n$ is necessary to take into account the
existence of a purely imaginary axial gravitational mode, but this is
just a matter of convention.

\subsubsection{Purely damped modes}
\label{PDSAdS}

The study of electromagnetic and gravitational quasi-normal modes for
S-AdS black holes has unexpectedly revealed the existence of axial
gravitational modes with pure-imaginary frequency \cite{CL}. These
modes have the interesting property that their frequency does not
scale linearly with the black hole radius, but rather with the {\it
inverse} of the black hole radius: an explicit fit (performed using
data in the range $r_+=10$ to $r_+=100$ at steps $\Delta r_+=0.1$)
shows indeed that for purely damped axial modes with $l=2$ (see Table
III in \cite{CL}) $\omega_{I}=1.335/r_+$.

In other words, in this case $\omega_I$ scales with the inverse of the
mass, as it does for Schwarzschild black holes in flat spacetime:
hence these modes are {\it particularly long lived}. Cardoso and Lemos
managed to prove stability for other kinds of perturbations, but not
for these modes. Indeed, the proof relies on the radial potential
being positive, while $V_{odd}$, given by formula (\ref{Vodd}), can be
negative for some values of $l$ and $M$.

Electromagnetic perturbations of large S-AdS black holes are also
characterized by a set of pure imaginary modes, which however scale in
the ``ordinary'' way ($\omega_I\sim r_+$).  

\begin{table}
\centering
\caption{
Lowest QNM of S-AdS, pure electromagnetic perturbations for $l=2$.  In
parentheses we give the values listed in Table II of the paper by
Cardoso and Lemos.
}
\vskip 12pt
\begin{tabular}{@{}ccc@{}}
$r_+$ &$\omega_R$~$(\omega_R^{CL})$ &$\omega_I$~$(\omega_I^{CL})$  \\
\hline
0.8  &3.224~(2.501) &0.996~(1.176)\\
1    &3.223~(2.496) &1.384~(1.579)\\
5    &3.090~(0.822) &9.822~(10.309)\\
10   &0	            &16.623~(15.755)\\
50   &0	            &75.269~(75.139)\\
100  &0             &150.133~(150.069)\\
\end{tabular}
\label{CLEM}
\end{table}

We point out that we get essentially perfect agreement with all the
numerical results shown in \cite{CL}, with the exception of their
Table II (lowest QNM of electromagnetic perturbations for $l=2$). The
values we obtain in this particular case are given in Table
\ref{CLEM}.  More importantly, we find no evidence for the axial
``algebraically special'' mode with $r_+=1$ and pure-imaginary
frequency $\omega_I=2$, indicated as a ``dubious'' result in Table III
of \cite{CL}. In both formulations of the eigenvalue problem, our
two-dimensional M\"uller root-searching routines fail to converge
there, and even a one-dimensional search on the axis $\omega_R=0$ does
not show any root. This is not too surprising. This algebraically
special mode (if it exists) would be characterized by boundary
conditions different from those of ordinary QNM's: there would be no
radiation going down the black hole horizon \cite{CL}.

It is well known from the asymptotically flat case \cite{C} that, as
we ``turn on'' the charge, electromagnetic and gravitational modes
become ``mixed'': as we noted earlier on, only when $Q\to 0$ the
potentials $V_2^{\pm}$ reduce to pure gravitational perturbations of
Schwarzschild, and the $V_1^{\pm}$ potentials reduce to pure
electromagnetic perturbations. We will see in the next section that
both modes (i.e., those which can be classified as ``purely
electromagnetic'' and ``pure gravitational'' in the S-AdS, $Q=0$
limit) show an interesting behaviour as the charge increases.  In
particular, {\it purely damped} modes of both classes behave in a very
peculiar way: their damping seems to go to infinity ($\omega_I\to 0$)
in the extremal black hole case, suggesting the possibility that {\it
extremally charged RN-AdS black holes may be marginally unstable to
electromagnetic and gravitational perturbations}.






\subsection{Reissner-Nordstr\"om-anti-de Sitter black holes}

\subsubsection{``Ordinary'' quasi-normal modes}

A first study of scalar perturbations of RN-AdS black holes was
carried out by Wang, Lin and Abdalla \cite{WLA}. The authors found
that ``switching on the charge'' results in a breakdown of the linear
relations $\omega_R(T)$, $\omega_I(T)$ (and in a similar breakdown of
the linear relation between $T$ and $r_+$) which is larger the larger
is the charge. In a following paper, time evolutions of the scalar
field wave equation were used to cross-check these results \cite{WMA}
and to study the late-time behaviour of the perturbing field. Both
numerical methods used in our work (see the Appendix) are
frequency-domain methods, and break down for large values of the
charge, for reasons related to the radius of convergence of
power-series solutions of the wave equation \cite{WLA}. However, time
evolutions do not suffer of this problem, and were used to show that
the imaginary part of the frequency attains a maximum at some critical
value of the charge $Q_{crit}$, to reduce again as $Q\to Q_{ext}$.
This is an important point on which we will comment later.

Our numerical results for the scalar case agree with those presented
in \cite{WLA,WMA} (for example, we are in excellent agreement with
Table II in \cite{WMA}), and confirm a breakdown of the linear
relations (\ref{linear2}) as the black hole charge increases. However,
in \cite{WLA} it was not only claimed that $\omega_I$ increases with
$Q$, but also that $\omega_R$ decreases with $Q$, so that, in their
words, {\it ``if we perturb a RN-AdS black hole with high charge, the
surrounding geometry will not ``ring'' as much and long as that of the
black hole with small $Q$''}. Our results do not confirm this statement.

This can be seen, for example, from figure \ref{fig1}. The continuous
wiggling ``tails'' departing from the scalar-mode line are numerical
solutions of the scalar eigenvalue problem for selected horizon radii
and increasing charge. From the plot it is clear that, as the charge
increases, the imaginary part of the frequency increases, according to
the predictions in \cite{WLA}; however, the real part of the frequency
does not show a monotonically decreasing behaviour!

\begin{table}
\centering
\caption{
Location of minima and maxima of $\omega_R(\bar Q)$ for the
fundamental mode.  Values listed correspond to some selected values of
the horizon radius $r_+$ and to different classes of perturbations.  A
dash (``-'') means that our numerical method fails to converge before
reaching the corresponding minimum or maximum in $\bar Q$.
}
\vskip 12pt
\begin{tabular}{@{}ccccccc@{}}
\multicolumn{1}{c}{$r_+$} 
&\multicolumn{2}{c}{Scalar ($l=0$)} 
&\multicolumn{2}{c}{Axial ($l=2$)} 
&\multicolumn{2}{c}{Polar ($l=2$)} \\
\hline
$r_+$  &min &max 
       &min &max
       &min &max
\\
\hline
100  & 0.366   & 0.474  & 0.366 & 0.474 & 0.366 & 0.474\\
50   & 0.366   & 0.474  & 0.366 & 0.474 & 0.366 & 0.474\\
10   & 0.367   & 0.475  & 0.368 & 0.476 & 0.369 & 0.477\\
5    & 0.372   & 0.480  & 0.376 & 0.483 & 0.376 & 0.488\\
1    & 0.468   & 0.571  & -     & -     & 0.503 & -    \\
\end{tabular}
\label{minmax}
\end{table}

\begin{figure}[h]
\centering
\includegraphics[angle=270,width=8.5cm,clip]{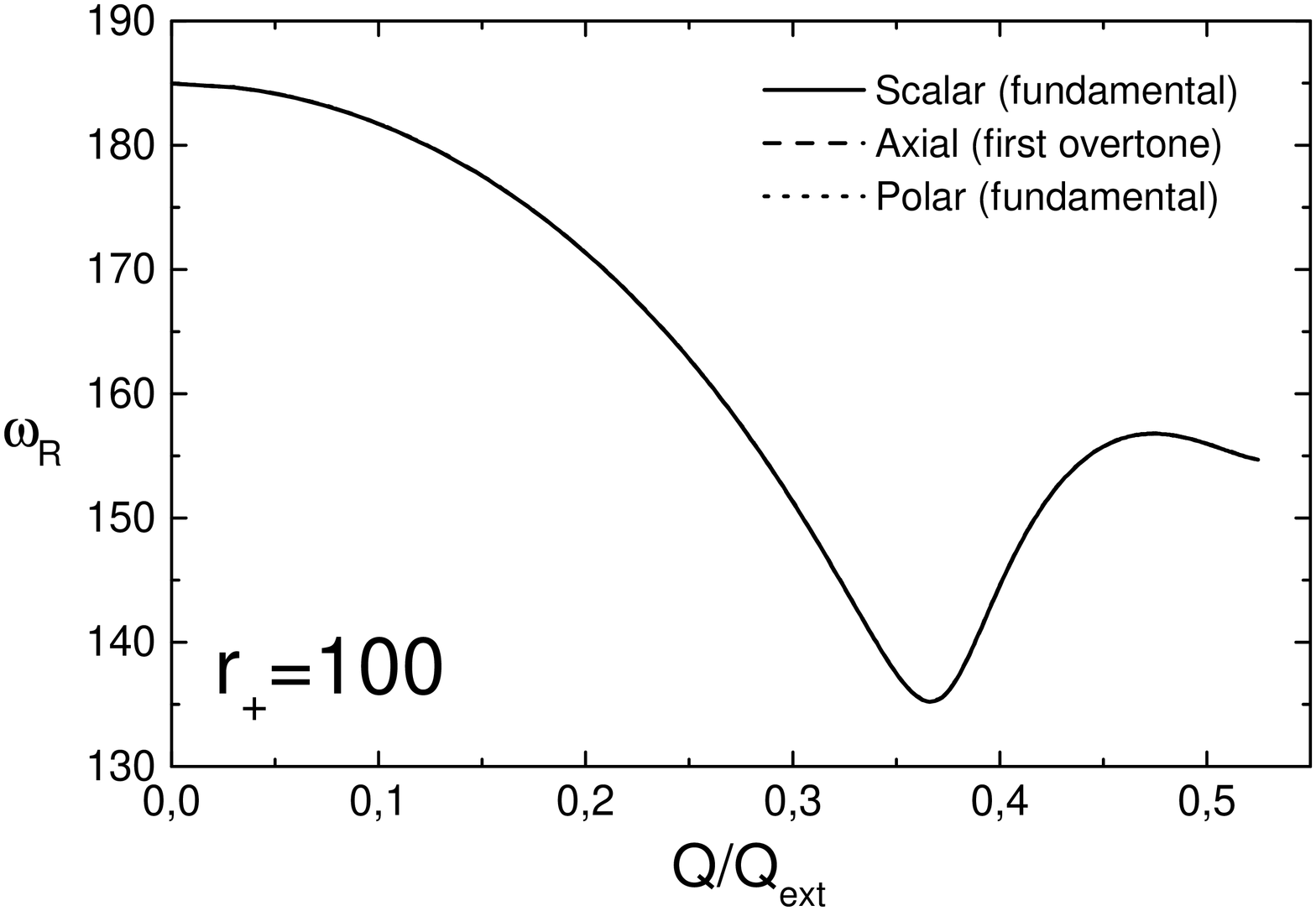}
\includegraphics[angle=270,width=8.5cm,clip]{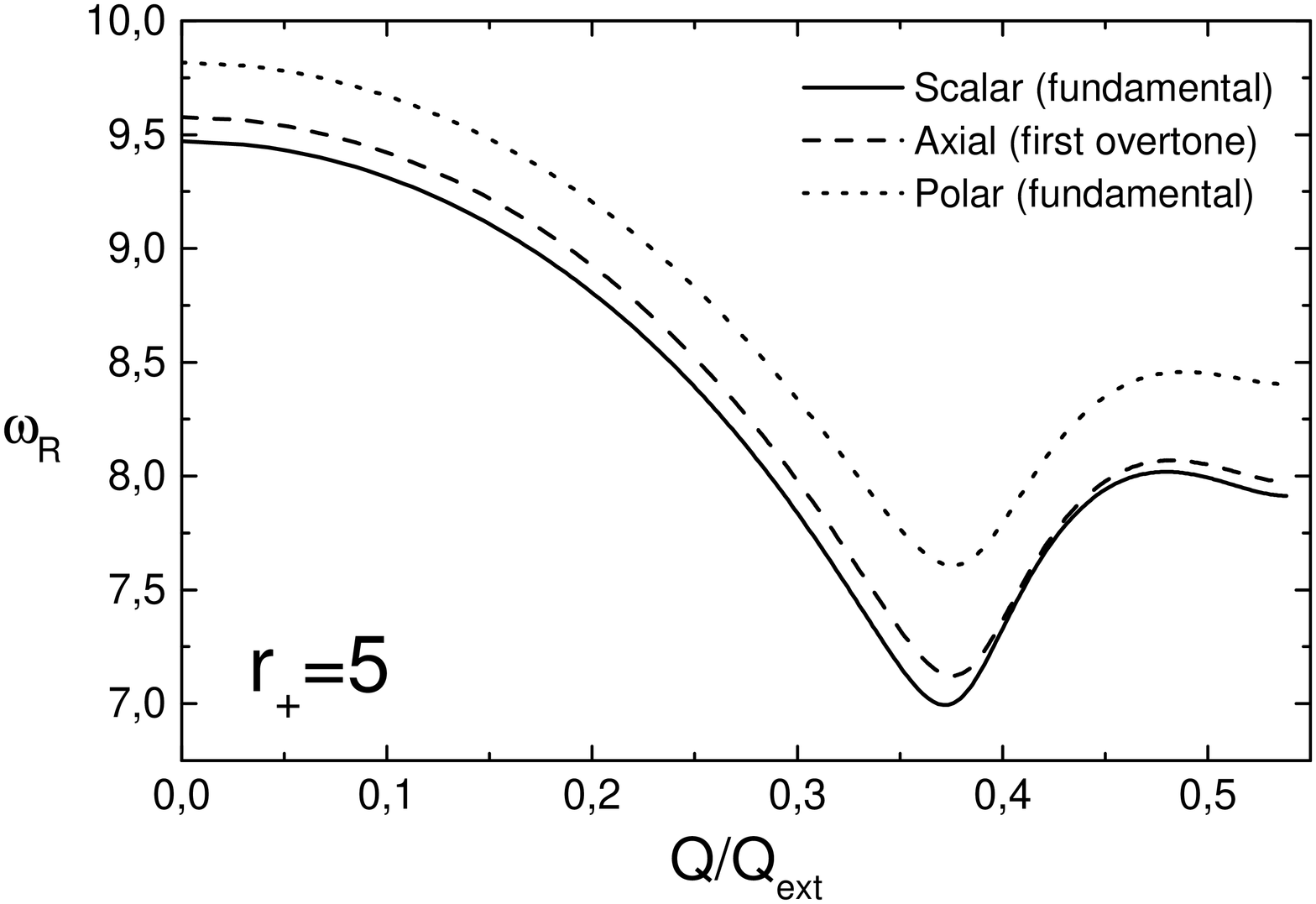}
\includegraphics[angle=270,width=8.5cm,clip]{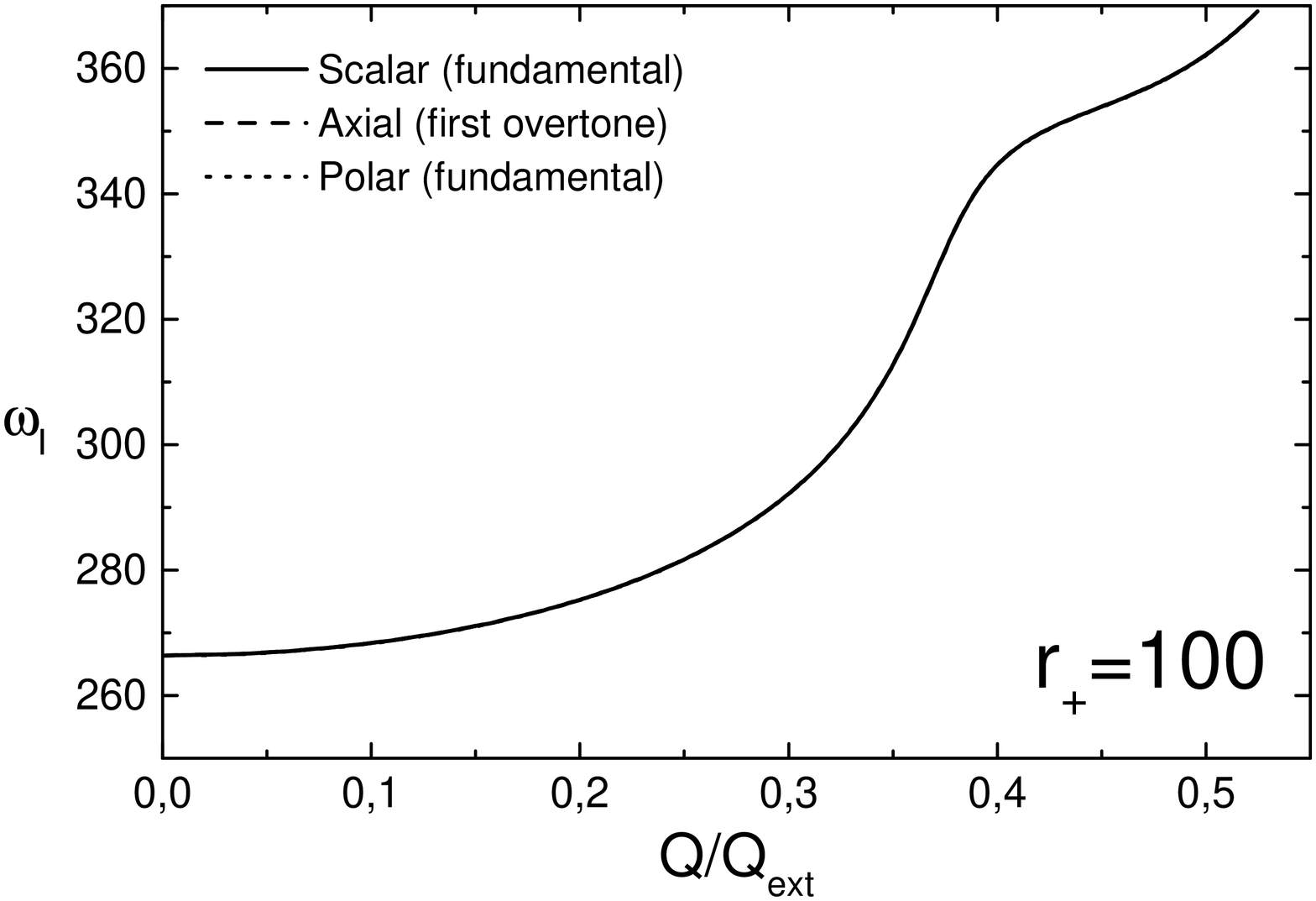}
\includegraphics[angle=270,width=8.5cm,clip]{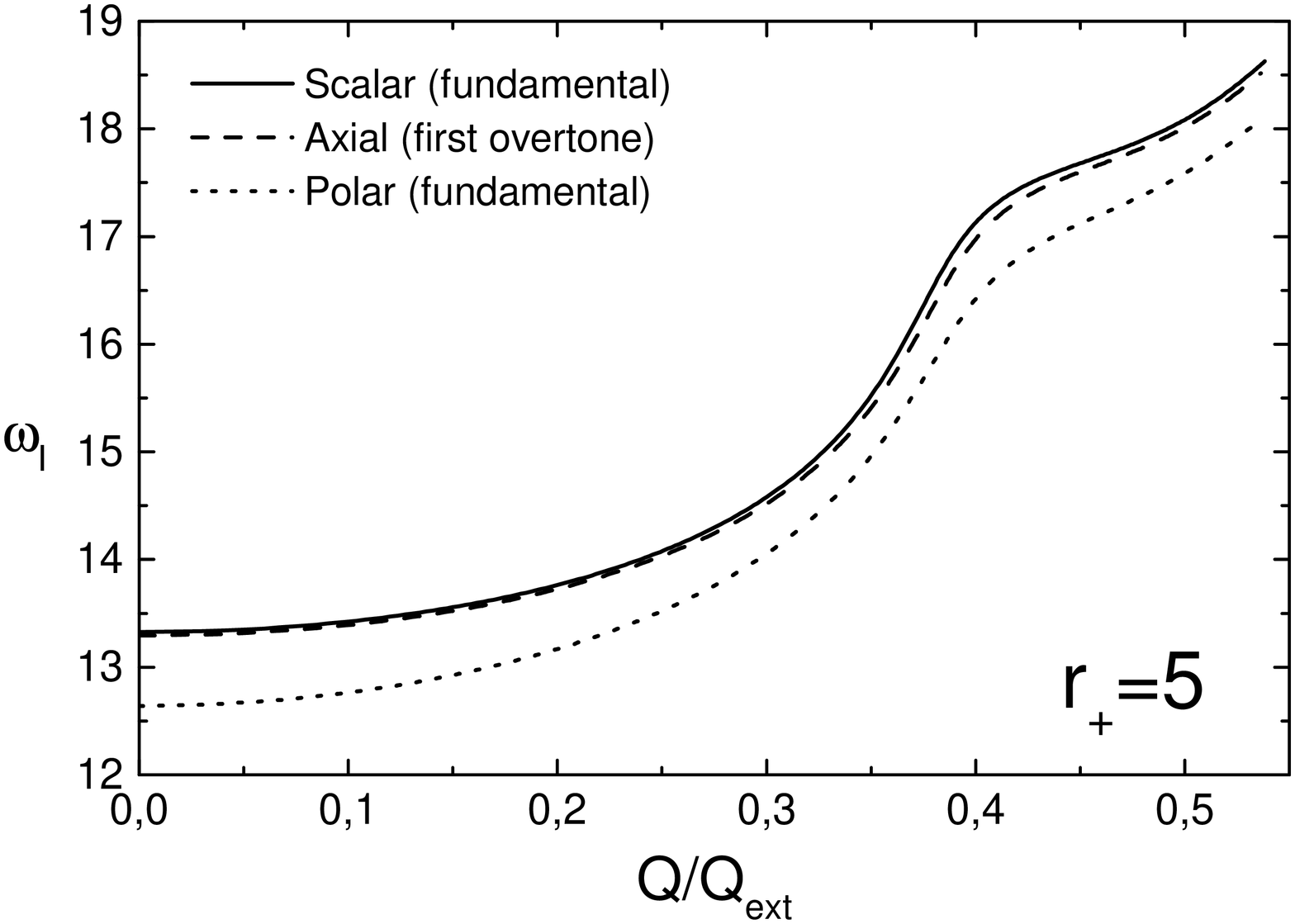}
\caption{
The top two panel shows the characteristic ``wiggling'' of the real
part of the QNM frequency, $\omega_R$, as a function of the normalized
charge $\bar Q=Q/Q_{ext}$, for selected values of the horizon radius
$r_+$; the bottom two panels show the imaginary part of the QNM
frequency, $\omega_I(\bar Q)$, for the same black holes.  Scalar
perturbations are denoted by solid lines, axial perturbations by
dashed lines, and polar perturbations by dotted lines.  The
isospectrality of different kinds of perturbations, which holds in the
large black hole limit, is clearly lost as the black hole ``size''
becomes comparable to the AdS radius, $r_+\sim 1$.  Notice that
$\omega_I''(\bar Q)=0$ when $\omega_R'(\bar Q)=0$ (a prime denoting
differentiation with respect to $\bar Q$).  
}
\label{fig2}
\end{figure}

The situation is clarified in figure \ref{fig2}, where we also
consider polar and axial perturbations.  Consider first the top two
panels in figure \ref{fig2}, where we plot the QNM frequency
$\omega_R$ as a function of the normalized charge $\bar
Q=Q/Q_{ext}$. Each panel corresponds to a different value of $r_+$. As
we could expect from the uncharged black hole limit, the scalar and
mixed gravitational/electromagnetic perturbations are {\it almost
isospectral for large black holes}, but the isospectrality breaks down
for intermediate-size holes. Quite interestingly, QNM frequencies show
indeed some kind of ``damped oscillations'' as $Q$ increases. In Table
\ref{minmax} we list the location of minima and maxima in
$\omega_R(\bar Q)$ for the three different kinds of perturbations, and
for selected values of $r_+$. For large black holes (which are of
greater interest in view of the AdS/CFT correspondence), the values of
$\bar Q$ for which the frequency attains the local minimum and maximum
are {\it independent of the black hole radius}: they correspond,
respectively, to $\bar Q=0.366$ and $\bar Q=0.474$.

For small values of the charge, $\bar Q<1/3$, an excellent fit of the
real part of the frequency at fixed $r_+$ (errors are typically of
order $10^{-4}$ or smaller, especially for the low-order modes), valid
for all kinds of perturbations, is provided by a simple polynomial
relation: 
\be\label{oRQ}
\omega_R=\omega_R^{(0)}\left(1-a\bar Q^2-b\bar Q^4\right),
\ee 
where we have denoted by $\omega_R^{(0)}$ the real part of the QNM
frequency in the zero-charge (S-AdS) limit.
The fit parameters $a$ and $b$ are given for selected values of $r_+$
in Table \ref{charge}. Again, they are largely independent of $r_+$
and of the kind of perturbation considered, at least in the large
black hole limit.

\begin{table}
\centering
\caption{
Fit parameters for the charge dependence of the real part of the QNM
frequency for some selected horizon radii and different classes of
perturbations -- see formula (\ref{oRQ}). The accuracy of these fits,
which are only valid for $\bar Q<1/3$, is typically of some parts in a
thousand.
}
\vskip 12pt
\begin{tabular}{@{}ccccccc@{}}
\multicolumn{1}{c}{$r_+$} 
&\multicolumn{2}{c}{Scalar ($l=0$)} 
&\multicolumn{2}{c}{Axial ($l=2$)} 
&\multicolumn{2}{c}{Polar ($l=2$)} \\
\hline
$r_+$  &$a$ &$b$ 
       &$a$ &$b$
       &$a$ &$b$
\\
\hline
100  & 1.6944   & 3.6144  & 1.6943   & 3.6135  & 1.6940   & 3.6107 \\
50   & 1.6939   & 3.6109  & 1.6935   & 3.6074  & 1.6923   & 3.5963 \\
10   & 1.6775   & 3.5021  & 1.6771   & 3.4198  & 1.6378   & 3.1729 \\
5    & 1.6284   & 3.1911  & 1.5882   & 2.9204  & 1.4695   & 2.2189 \\
\end{tabular}
\label{charge}
\end{table}

Let us now turn to the bottom two panels in figure \ref{fig2}.  For
the values of the charge allowed by our numerical methods, $\omega_I$
increases monotonically as $\bar Q$ increases.  The second derivative
of $\omega_I(\bar Q)$ changes sign when the real part of the frequency
attains the local minimum, and then again at the local maximum .  A
good fit in the region $\bar Q<1/3$ (errors being of the order of a
few percent for the fundamental mode, and of order 1 \% or less for
the overtones) is again obtained using a polynomial relation:
\be\label{oIQ}
\omega_I=\omega_I^{(0)}\left(1+c\bar Q^2+d\bar Q^4\right),
\ee
where we denote by $\omega_I^{(0)}$ the imaginary part of the QNM
frequency in the S-AdS limit.
The fit parameters $c$ and $d$ are listed in Table \ref{chargeI}.  As
for the fitting coefficients of the real part, $c$ and $d$ are largely
independent of $r_+$ and of the kind of perturbation considered in the
large black hole limit.

\begin{table}
\centering
\caption{
Fit parameters for the charge dependence of the imaginary part of the QNM
frequency for some selected horizon radii and different classes of
perturbations -- see formula (\ref{oIQ}). The accuracy of these fits,
which are only valid for $\bar Q<1/3$, is typically of some parts in a
hundred.
}
\vskip 12pt
\begin{tabular}{@{}ccccccc@{}}
\multicolumn{1}{c}{$r_+$} 
&\multicolumn{2}{c}{Scalar ($l=0$)} 
&\multicolumn{2}{c}{Axial ($l=2$)} 
&\multicolumn{2}{c}{Polar ($l=2$)} \\
\hline
$r_+$  &$c$ &$d$ 
       &$c$ &$d$
       &$c$ &$d$
\\
\hline
100  & 0.56827   & 5.8567  & 0.56829   & 5.8556  & 0.56892   & 5.8550   \\
50   & 0.56839   & 5.8524  & 0.56847   & 5.8483  & 0.57096   & 5.8458   \\
10   & 0.57193   & 5.7182  & 0.57378   & 5.6222  & 0.63597   & 5.5560   \\
5    & 0.58145   & 5.3361  & 0.58634   & 5.0124  & 0.83131   & 4.7163   \\
\end{tabular}
\label{chargeI}
\end{table}

Summarizing, for large black holes with charge $\bar Q<1/3$, QNM
frequencies and dampings are approximated within about 1 \% by a very
simple formula:
\beq\label{RNAdSfit}
\omega_R&=&(7.75+5.46n)T^{(0)}\left(1-1.694\bar Q^2-3.61\bar Q^4\right),\\
\omega_I&=&(11.16+9.43n)T^{(0)}\left(1+0.568\bar Q^2+5.86\bar Q^4\right),\nn
\eeq
where $T^{(0)}=(1+3r_+^2)/4\pi r_+$ is the S-AdS (uncharged) black
hole temperature.  Notice that, for fixed $Q$, the dependence of
$\omega_R$ and $\omega_I$ on the angular index $l$ is extremely weak,
so this formula holds with good accuracy also for angular multipoles
larger than the fundamental.

\subsubsection{Purely damped modes: are extreme RN-AdS black holes marginally 
unstable?}

Modes which become ``purely electromagnetic'' or purely imaginary in
the S-AdS limit deserve a separate discussion.

\begin{figure}[h]
\centering
\includegraphics[angle=270,width=8.5cm,clip]{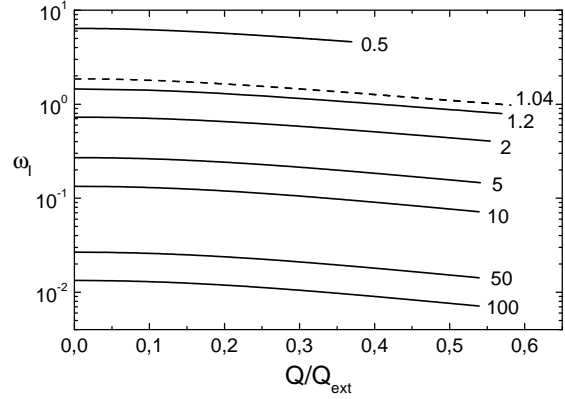}
\caption{
Imaginary part of the purely damped mode reducing to pure axial
gravitational perturbations with $l=2$ in the zero-charge
limit. Starting from S-AdS results, we track the modes for selected
values of the horizon radius $r_+$ (indicated to the right of each
curve). The calculation is terminated when our root finder fails to
converge. The dashed line indicates the last mode we can find before
the numerical method breaks down as we approach the ``algebraically
special'' ($\omega_I=2$, $r_+=1$) frequency.
}
\label{fig3}
\end{figure}

Let us first consider the pure-imaginary modes for axial gravitational
perturbations. In figure \ref{fig3} we ``track'' the imaginary part of
these modes, starting from the corresponding zero-charge limit, for
different values of the horizon radius. As explained when we discussed
purely imaginary QNM's for S-AdS black holes, our numerical results do
not confirm the existence of an algebraically special mode in the
zero-charge limit at $r_+=1$: as we decrease $r_+$, the root finder
breaks down when $r_+\sim 1.04$ (dashed line in the plot),
corresponding to a S-AdS frequency $\omega_I=1.860$. 

It is clear from the plot that all modes show a tendency to approach
the horizontal axis as the charge approaches the extremal value. This
tendency is much clearer if we use a linear scale (instead of the log
scale used in this plot, which is necessary to display all modes
together). Unfortunately, for the convergence reasons discussed
earlier, we cannot push our calculations to the extremal
limit. However, our plots lead us to conjecture that {\it all
pure-imaginary axial modes are such that $\omega_I\to 0$ as $Q\to
Q_{ext}$}.

\begin{figure}[h]
\centering
\includegraphics[angle=270,width=8.5cm,clip]{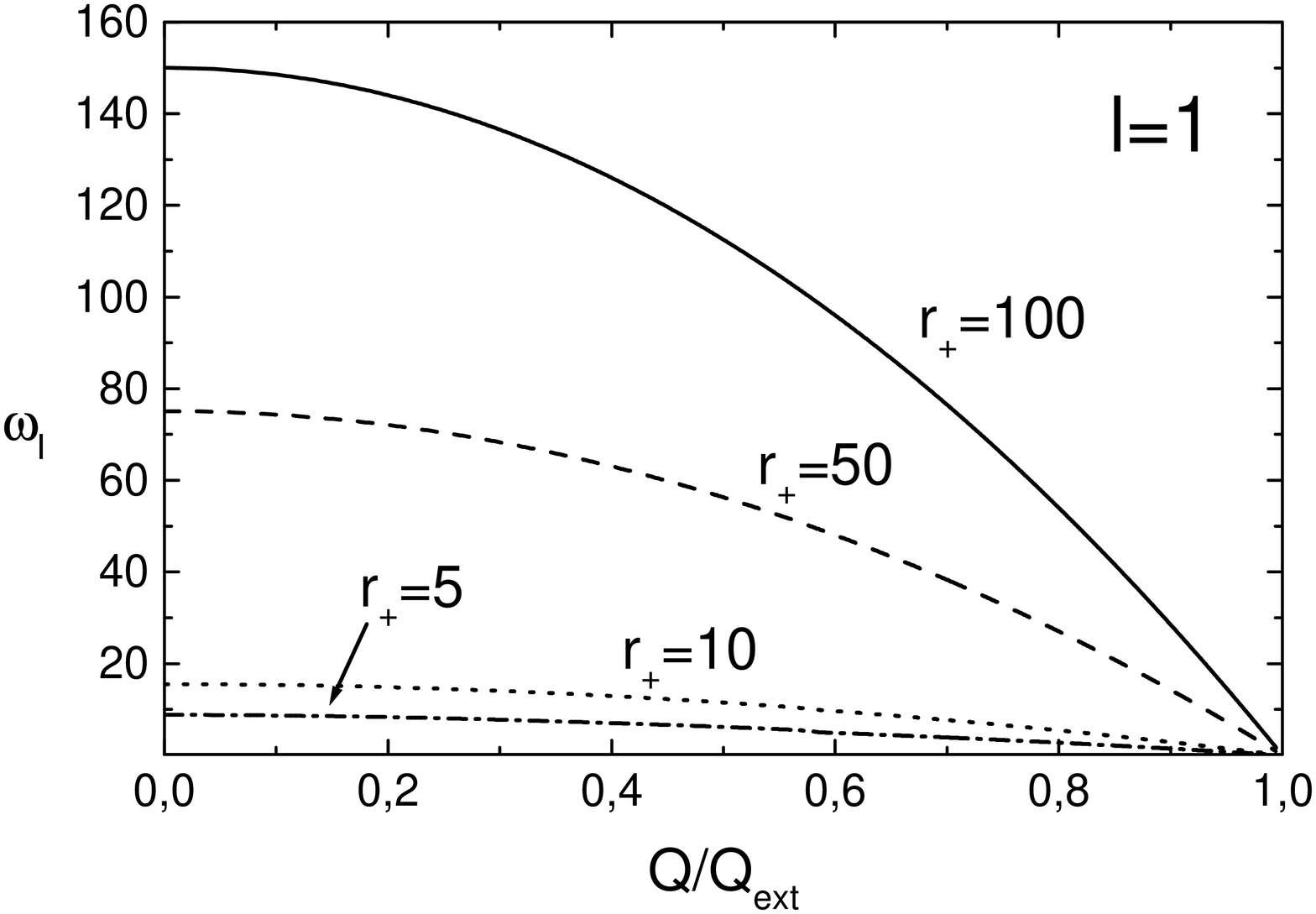}
\includegraphics[angle=270,width=8.5cm,clip]{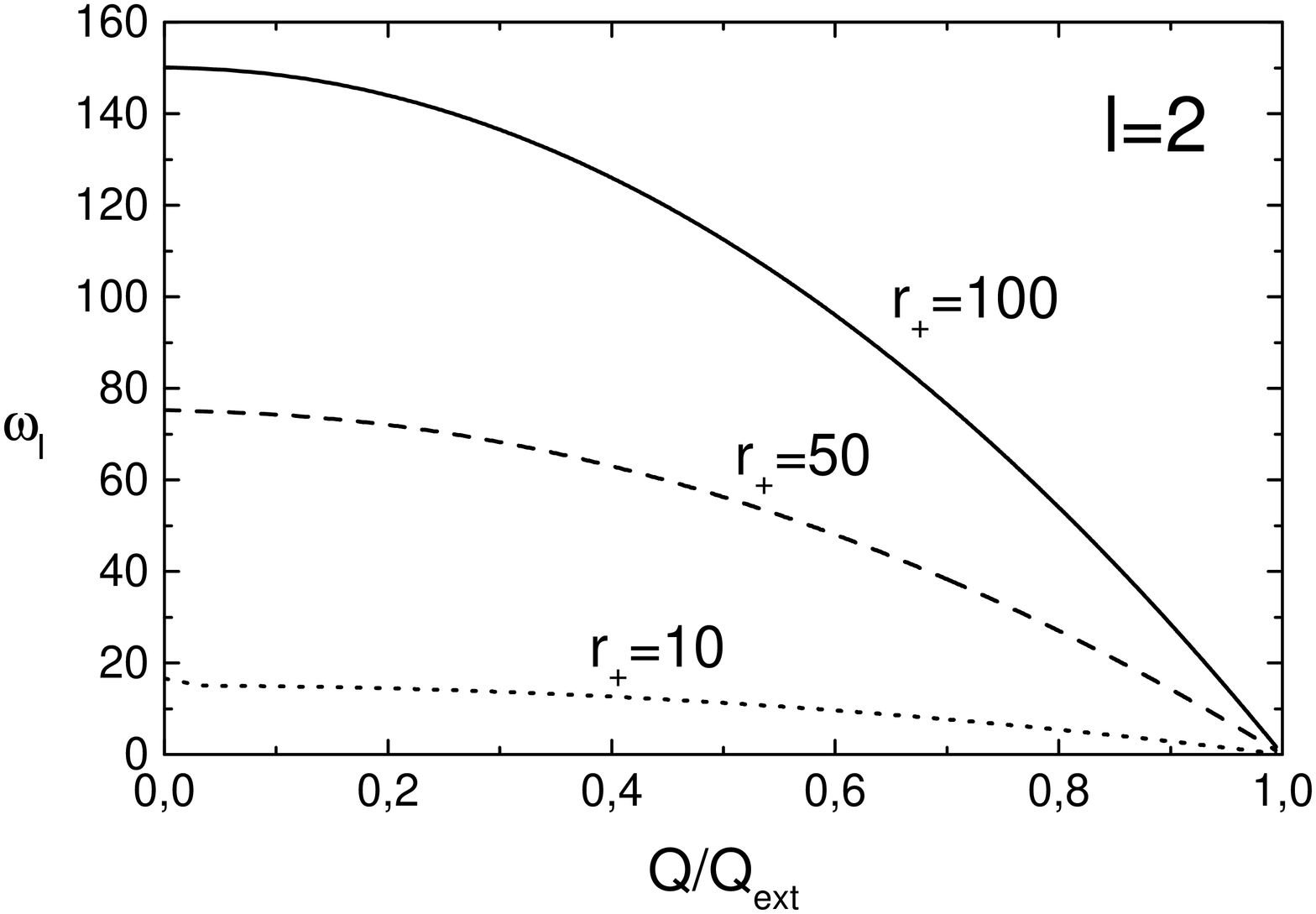}
\caption{
Imaginary part of the purely damped mode reducing to pure
electromagnetic perturbations with $l=1$ (left panel) and $l=2$ (right
panel) in the zero-charge limit. Continuous, dashed and dotted lines
refer, respectively, to black holes having horizon radius
$r_+=100,~50$ and $10$ (top to bottom in the plots). For $l=1$, we also
plot the pure imaginary mode with $r_+=5$ (dash-dotted line).
}
\label{fig4}
\end{figure}

Our conjecture finds firmer ground in the behaviour of the modes for
the $V_1^-$ potential which reduce to purely-damped electromagnetic
perturbations with $l=1$ and $l=2$ in the S-AdS limit. In this case,
we were able to track roots for all values of the charge $\bar Q<1$,
and our results are shown in the two panels of figure \ref{fig4}: the
imaginary part of purely-electromagnetic modes in the Schwarzschild
limit {\it does} indeed tend to zero as $\bar Q\to 1$.

\begin{figure}[h]
\centering
\includegraphics[angle=270,width=8.5cm,clip]{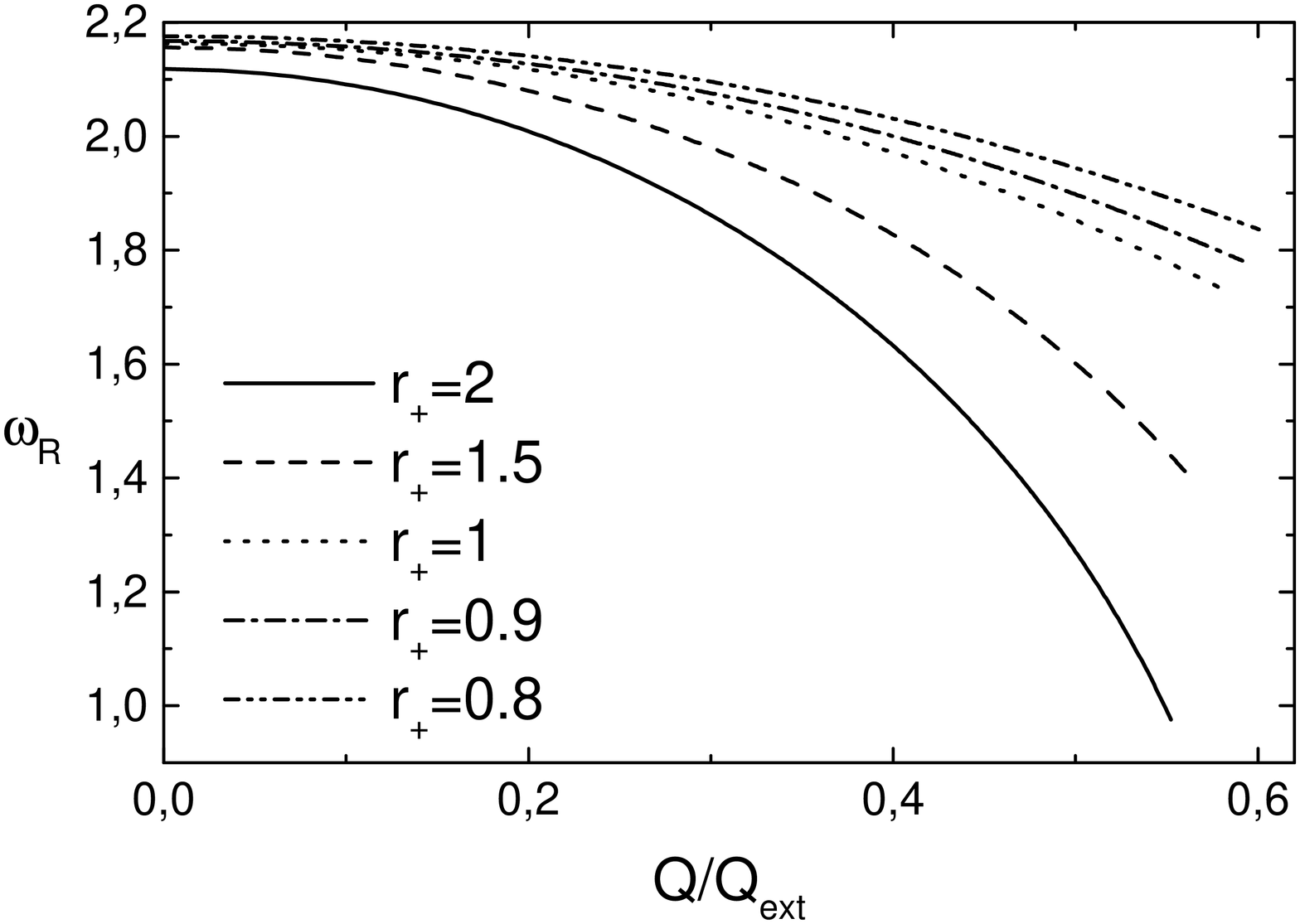}
\includegraphics[angle=270,width=8.5cm,clip]{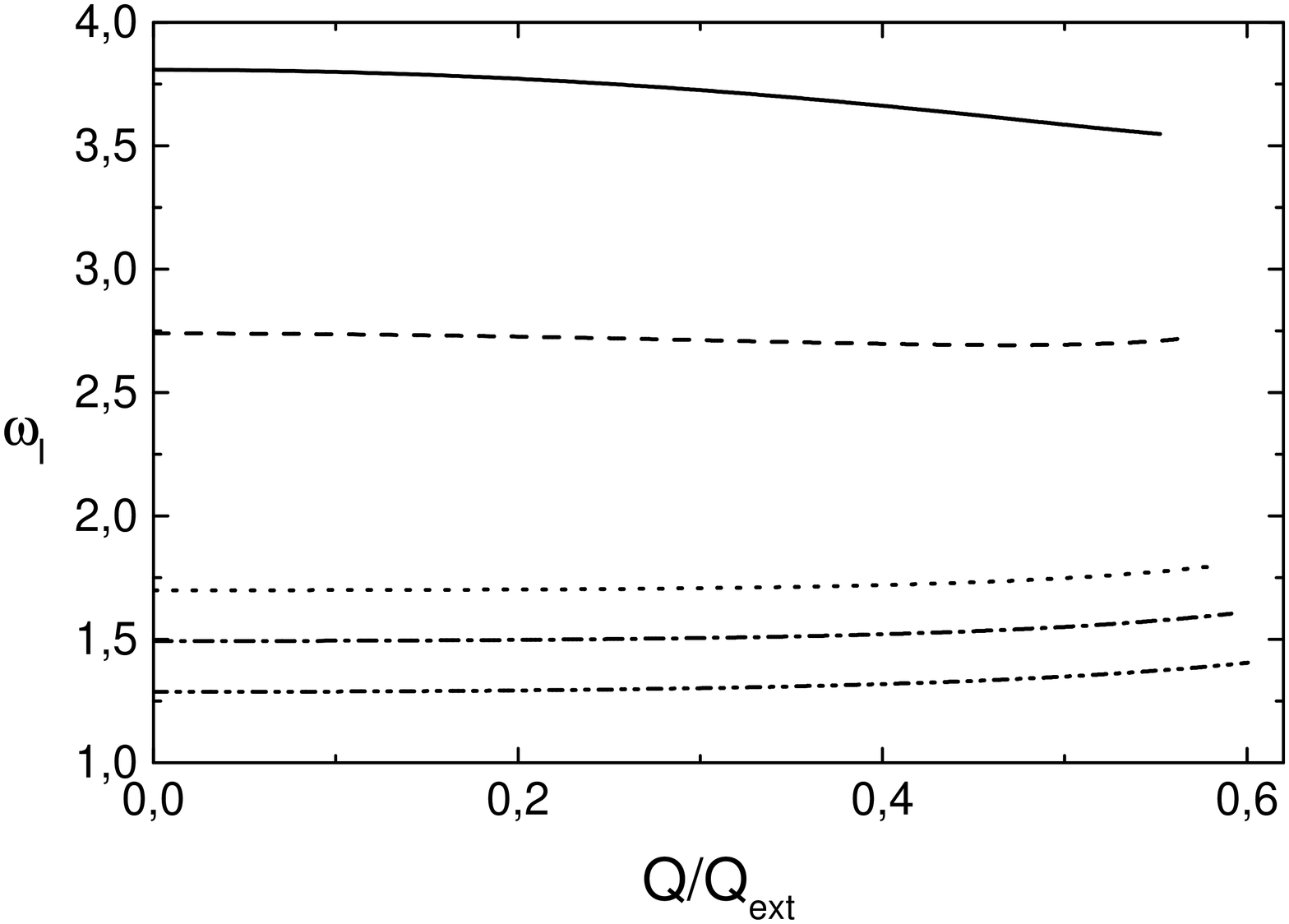}
\caption{
Real (left panel) and imaginary parts (right panel) of the small-black
hole QNM frequencies for modes reducing to pure electromagnetic
perturbations with $l=1$ in the zero-charge limit. The size of the
black hole horizon radius corresponding to each curve is indicated in
the inlay.
}
\label{fig5}
\end{figure}

For small black holes, such a behaviour is not observed. Small black
holes in S-AdS do not have pure imaginary electromagnetic
modes. Tracking again the modes of the $V_1^-$ potential (figure
\ref{fig5}) we see that their real part decreases as a function of
$\bar Q$, but the imaginary part does not: indeed, $\omega_I(\bar Q)$
is decreasing when $r_+\gtrsim 2$, but it shows a relative minimum
when $r_+=1.5$, and is roughly constant or slowly increasing when
$r_+\lesssim 1$!

Our results suggest the possibility that, {\it unless the mode
amplitude tends to zero in the extremal limit, large extremal RN-AdS
black holes are marginally unstable to electromagnetic (and perhaps,
axial gravitational) perturbations}. This result may be extremely
interesting in view of the AdS/CFT conjecture.

We only make a couple of comments at this stage. First, notice that
even for S-AdS black holes, a proof of stability to large black hole,
axial gravitational perturbations does not exist \cite{CL}.  Second,
as we recalled earlier, time evolutions of scalar fields in RN-AdS
spacetimes have shown that, for $Q$ greater than some critical value
$Q_{crit}$, $\omega_I$ is a {\it decreasing} function of charge
\cite{WMA}. The authors suggested a possible connection between this
behaviour and the existence of a second-order phase transition for
extremal RN-AdS black holes \cite{CEJM}. Development of more efficient
numerical methods is needed to verify such a connection between QNM's
and thermodynamical phase transitions of extremal RN-AdS black
holes. An understanding of the link between our study and the
dynamical instability discussed by Gubser and Mitra \cite{GM1,GM2},
which in their words persists {\it ``for finite size black holes in
AdS, down to horizon radii on the order of the AdS radius''}, could
shed further light on the AdS/CFT conjecture. We notice that an
interesting discussion of the connection between dynamical and
thermodynamical stability in AdS spacetimes can be found in \cite{HR}.

\section{Conclusions}

We have studied scalar, electromagnetic and gravitational
perturbations of a RN-AdS spacetime, and computed its QNM's. Our
results extend previous computations carried out for scalar,
electromagnetic and gravitational perturbations of S-AdS black holes
\cite{HH,CL} and for scalar perturbations of RN-AdS black holes
\cite{WLA}. For ``large'' black holes, whose horizon is much larger
than the AdS radius, different kinds of perturbations are almost
exactly {\it isospectral}; this isospectrality is broken when the
black hole's horizon radius is comparable to the AdS radius. Contrary
to previous claims, we find that, as the black hole charge $Q$
increases, the real part of the QNM frequency does not continuously
decrease, but rather shows a minimum followed by a maximum. The
location of these extrema as a function of $\bar Q=Q/Q_{ext}$ is
independent of the black hole radius for large black holes: minima are
located at $\bar Q=0.366$, and maxima at $\bar Q=0.474$.  We found
that the imaginary part of the QNM frequency $\omega_I$ is generally,
at least for small values of $Q$, an increasing function of the black
hole charge, whose second derivative changes sign corresponding to the
minima and maxima of $\omega_R(\bar Q)$.  We also obtained the very
accurate fitting formulas (\ref{RNAdSfit}), which are valid for large
black holes having charge $Q<Q_{ext}/3$.

A very interesting result concerns electromagnetic and axial
perturbations of large black holes. These perturbations are
characterized by the existence of purely damped modes. We advanced
numerical evidence that the damping of these modes approaches infinity
as the black hole charge approaches the extremal value. If the
corresponding mode amplitude does not tend to zero, this would imply
that {\it extremally charged RN-AdS black holes are marginally
unstable}.  Our results are relevant in view of the AdS/CFT
conjecture, since, according to it, the AdS QNM's yield thermalization
timescales in the corresponding conformal field theory. It would be
desirable to relate our results to the existence of a second-order
phase transition for extremal RN-AdS black holes \cite{CEJM} and to a
dynamical instability which has been studied in the past
\cite{GM1,GM2}, and which also holds for finite size RN-AdS black
holes having horizon radii larger than the AdS radius.  We believe
that investigating these connections could clarify many issues related
to the AdS/CFT correspondence.

\acknowledgments
This work has been supported by the EU Programme 'Improving the Human
Research Potential and the Socio-Economic Knowledge Base' (Research
Training Network Contract HPRN-CT-2000-00137).

\appendix
\section{Numerical computation of quasinormal modes}
\subsection{The Horowitz--Hubeny method}
\label{nummeth}
Here we briefly describe the first numerical method we used to compute
the quasinormal modes. This method has been described elsewhere
\cite{HH,CL,WLA}, so we only give the essential formulas, which
encompass the slightly more general case considered here.  For a
generic wavefunction obeying one of the five wave equations
(\ref{scalar}), (\ref{axial}) or (\ref{polar}), write

\be
\psi(r)=e^{i\omega r_*}Z(r)
\ee
and get the equation for $\psi(r)$
\be
f(r){\p^2 \psi(r)\over \p r^2}
+\left[f'(r)-2i\omega\right]{\p \psi(r)\over \p r}
-{\tilde V(r)}\psi(r)=0,
\ee
where ${\tilde V(r)}=V(r)/f(r)$. Introducing a new variable $x=1/r$
to ``compactify'' the region outside the black hole, and defining
$x_+=1/r_+$, this equation is rewritten as
\be\label{wavex}
s(x){d^2 \psi(x)\over dx^2}
+{t(x)\over x-x_+}{d \psi(x)\over dx}
+{u(x)\over (x-x_+)^2}\psi(x)=0
\ee
where
\beq
s(x)&=&{
r_0x^5-x^4-x^2-Q^2x^6\over x-x_+
},\\
t(x)&=&3r_0x^4-2x^3-4Q^2x^5-2i\omega x^2,\\
u(x)&=&(x-x_+)\tilde V(x),
\eeq
and
\be
r_0={1+x_+^2+Q^2x_+^4\over x_+^3}.
\ee
Now look for a series solution in the form
\be
\psi(x)=\sum_{k=0}^\infty a_k(x-x_+)^k.
\ee
This series solution obeys the boundary conditions required for
quasinormal modes if the condition
\be
\psi(0)=\sum_{k=0}^\infty a_k(-x_+)^k=0
\ee
is satisfied. The coefficients $a_k=a_k(\omega)$ can be obtained
through the following recursion relation:
\be
a_j=-{1\over P_j}\sum_{k=0}^{j-1}\left[
k(k-1)s_{j-k}+kt_{j-k}+u_{j-k}
\right]a_k.
\ee
In this formula the $s_k$'s, $t_k$'s and $u_k$'s are coefficients in a
Taylor series expansion around $x=x_+$ of the functions $s(x)$, $t(x)$
and $u(x)$ respectively, and
\be
P_j=j(j-1)s_0+jt_0.
\ee

\subsection{The Fr\"obenius method}
\label{nummeth2}

As a check of our results, we have applied an alternative method,
originally suggested by Leaver \cite{L} and applied in \cite{MN} to
the axial perturbations of S-AdS BHs. We have generalized the approach
to RN-AdS BHs. Since this method is computationally much faster, and
very easy to implement for the axial potentials $V_{1}^-$ and $V_2^-$,
we give the details here.

Rewrite equation (\ref{wavex}) multiplying it by $(x-x_+)/x^2$ to find:
\be
p(x){d^2 \psi(x)\over dx^2}
+q(x){d \psi(x)\over dx}
+{\tilde V(x)\over x^2}\psi(x)=0
\ee
where as usual $\tilde V=V/f$, and furthermore
\beq
p(x)&=&(2Mx^3-x^2-1-Q^2x^4),\nn\\
q(x)&=&(6Mx^2-2x-4Q^2x^3-2i\omega).
\eeq

Now look for the solution in the form of a Fr\"obenius series which is
regular at the event horizon:
\be
u(x)=\sum_{n=0}^\infty a_n\left({x-x_+\over -x_+}\right)^n,
\ee
where $a_0=1$.

In the RN-AdS case, as is the case for Reissner-Nordstr\"om black
holes in asymptotically flat spacetimes \cite{L2}, the expansion
coefficients are determined by a four-term recurrence relation of the
form:
\beq
&&\alpha_0 a_1+\beta_0 a_0=0,\nn\\
&&\alpha_1 a_2+\beta_1 a_1+\gamma_1 a_0=0,\nn\\
&&\alpha_n a_{n+1}+\beta_n a_n+\gamma_n a_{n-1}+\delta_n a_{n-2}=0,
\qquad n=2,3,\dots
\eeq
where the recursion coefficients are given in terms of the parameters
appearing in the black hole potentials $V_i^-$, defined in equation
(\ref{axpot}), as follows:
\beq
&&\alpha_n=2(n+1)\left[(n+1)(1-3Mx_++2Q^2x_+^2)+i\omega/x_+\right],\nn\\
&&\beta_n=n(n+1)\left(6Mx_+-1-6Q^2x_+^2\right)
+l(l+1)-p_jx_+-{p_ip_j\over 2n}x_+^2,\nn\\
&&\gamma_n=(n^2-1)\left(4Q^2x_+^2-2Mx_+\right)
+p_jx_++{p_ip_j\over n}x_+^2,\nn\\
&&\delta_n=-(n-2)(n+1)Q^2x_+^2-{p_ip_j\over 2n}x_+^2.
\eeq

In the zero-charge limit $Q=0$ we recover a standard, three term
recursion relation \cite{MN}.  Once the $a_n$'s are known (in
practice, up to some finite value of $n$ determined by the accuracy we
want to achieve), the QNM's are determined by imposing the condition
$u(x=0)=0$.


\end{document}